# The Role of Solar Soft X-rays Irradiance in Thermospheric Structure

Srimoyee Samaddar[1], Karthik Venkataramani [1], Scott. M. Bailey [1]

[1] Center for Space Science and Engineering and Bradley Department of Electrical and Computer Engineering, Blacksburg, Virginia, USA.

Corresponding Author: Srimoyee Samaddar (srimoy1@vt.edu)

**Key Points**

1. We use a new one dimensional global average model, ACE1d, to show that ionization energy deposited in the lower thermosphere at 100-150 km, affects the temperature of the entire thermosphere

2. ACE1d model is used to demonstrate that the mechanism of temperature change is due to the molecular diffusion by non-thermal electrons.

3. Uncertainty in the modeled soft x-rays irradiance results in uncertainties in modeled thermospheric temperature.

**Abstract**

We use a new Atmospheric Chemistry and Energetics one-dimensional (ACE1D) thermospheric model to show that the energies deposited by the solar soft x-rays in the lower thermosphere at altitudes between 100 -150 km (Bailey et al. 2000), affects the temperature of the entire Earth's thermosphere even at altitudes well above 300 km.

By turning off the input solar flux in the different wavelength bins of the model iteratively, we are able to demonstrate that the maximum change in exospheric temperature is due to the changes in the soft x-ray solar bins.

We also show, using the thermodynamic heat equation, that the molecular diffusion via non-thermal photoelectrons, is the main source of heat transfer to the upper ionosphere/thermosphere and results in the increase of the temperature of the neutral atmosphere. Moreover, these temperature change and heating effects of the solar soft x-rays are comparable to that of the strong HeII 30.4nm emission.

Lastly, we show that the uncertainties in the solar flux irradiance at these soft x-rays wavelengths result in corresponding uncertainties in modeled exospheric temperature and the uncertainties increase with increased solar activity.

**General Language Summary**

Solar soft x-rays flux below 30.4 nm are highly energetic particles that can have detrimental effects on humans were they not easily absorbed by our atmosphere. However, the absorption of soft x-rays by the Earth's atmosphere above 100 km creates ionized and energized particles. These energetic changes can affect and even damage the satellites in Earth's low earth orbit, cause radio communication blackouts and radiation storms. Therefore, we need to have good models that can quantify these changes in order to correctly predict their effects on our atmosphere and consequently mitigate any harmful effects.

In this study, we use the one-dimensional global average ACE1D model to show that the soft x-rays deposit energy in the lower thermosphere at altitudes between 100 -150 km and how this energy is conducted by non-thermal photoelectrons created by ionization and leads to the increase in temperature of the entire Earth's thermosphere.

Lastly, we show that the uncertainties in the solar flux irradiance at these shorter wavelengths result in corresponding uncertainties in the modeled exospheric temperature.

1. Introduction

The solar soft x-ray region shortward of 30 nm, is a highly variable region of the solar spectra. It covers energy ranges over four orders of magnitudes and the solar irradiance varies over ten orders of magnitudes. This region consists of highly energetic flux with 10s of keV energies and the flux changes with change in activities of the sun. These highly energetic photon fluxes deposit their energy in the lower thermosphere at altitudes between 100- 150 km in then E region of the ionosphere, which causes the photoionization of the major neutral constituents $N_2$, $O_2$ and O, in the atmosphere and the formation of energetic photoelectrons. These energetic particles can cause further ionization or excitation or lead to heating of the atmosphere through collisions with other neutrals, ions or electrons. The huge energy dump from the soft x-rays disturbs the Earth's ionosphere, which then sets to readjust the diffusive equilibrium by a number of complicated heating and cooling processes, can be studied by analyzing the changes in its neutral temperature.

The wavelengths affecting these energetic processes lie mainly in the X-ray and Extreme UltraViolet (XUV) which is below 30 nm. Unfortunately, the data available for the soft x-ray irradiance is highly inconsistent which leads to subsequent inconsistencies in the modeled neutral temperature.

Soft x-rays data are generally available from different experiments carried out via satellites and rockets as early as the 1960s. Summary of the various instruments can be found in Woods et al. [2004]. Early measurements by instruments aboard the Atmosphere Explorer satellites and rocket measurements (Hinteregger et al. [1981]), SOLRAD (Kreplin [1970]; Kreplin and Horan[1992]), have led to several reference spectra in the EUV wavelengths. The uncertainty in these data is 20% to 40%. There are more recent spacecraft measurements in the X-ray and XUV regions such as EVE SAM and ESP data (0.1- 105 nm ) (Woods et al. [2010]), TIMED SEE (0.1-35 nm) (Woods et al. [2000]) , SNOE SEM data (2-7 nm, 6-19 nm and 17- 20 nm) (Bailey et al. [2000])which reduces this uncertainty. SNOE SEM data has reduced the uncertainty in the soft x-rays irradiance to a factor of about two. However, all these measurements are broadband which do not resolve the high variabilities in the soft x-rays range. New measurements like the X123 X-ray spectrometer on the Miniature X-ray Solar Spectrometer (MinXSS), that has an energy range of 0.5–30 keV with a nominal 0.15 keV energy resolution (Woods et al. 2017), are trying to resolve this issue.

When the data for the solar irradiance is scarce, we use empirical and physical models. There are different models available which can replicate the solar flux during different active and quiet conditions of the sun. The HFG model uses the SC21REFW reference spectra (Hinteregger [1981 ]) and a two-class (chromospheric or coronal) contrast ratio method to scale the reference spectrum. The EUVAC model also uses the SC21REFW reference spectra (Hinteregger [1981]) and the F10.7 index as inputs to an empirical model to obtain the solar flux for different solar activity conditions, the details of which are described later. The Flare Irradiance Spectral Model (FISM) (Chamberlin et al [2008]) is another empirical model which is used particularly to predict the flare spectrum from 0.1 to 190 nm at 1 nm resolution with a time cadence of 60 s, using a number of flare data sets from the instruments XPS (XUV Photometer System) and EGS (EUV Grating Spectrograph) of the TIMED SEE , SORCE XPS and also Geostationary Operational Environmental Satellite X-Rays Sensors (GOES XRS).

In order to study the effects of the soft x-rays on Earth's thermospheric temperature, we use a modified version of the EUVAC spectrum as an input to a new one-dimensional thermospheric model called Atmospheric Chemistry and Energetics (ACE1D). The ACE1D is a global average thermosphere-ionospheric model that self-consistently solves the continuity and energy equations to output the global mean neutral, ion and electron temperatures and densities. A detailed description of the model can be found in Venkataramani [2018]. A few important concepts such as the solar irradiance input, the neutral temperature calculation and the heating mechanisms are restated here for the purpose of explaining the physics behind the results of this paper.

## 1.1 Solar Fluxes

The input solar reference spectrum consists of 37 wavelength bins from 0.05-175 nm. The first 22 bins from 0.05-105 nm is based on the solar spectral model as proposed by Solomon and Qian [2005] and consists of scaling and rebinning the high resolution SC21REFW reference spectra (Hinteregger [1981]) into EUVAC proxy model (Richards et al. [1994]). The last fifteen 5nm bins are obtained from work of Woods and Rottman [2002] and are used to represent fluxes between 105 -175 nm. The solar spectrum for a given level of solar activity is calculated using the EUVAC model (Richards et al. [1994]) and described as follows:

$$F(\lambda) = F_{ref}(\lambda)[1 + A(\lambda)(P - 80)] \tag{1a}$$

where $\lambda$ is the wavelength, $F_{ref}$ is the reference solar spectrum and A is a wavelength dependent scaling factor. P is a proxy for solar activity, given by P=(F10.7_d+F10.7_a)/2 where F10.7_d and F10.7_a are the daily and the centered 81-day averages of 10.7 cm solar flux index for P ≥ 80.

For F<80, the fluxes are scaled to 80% of their reference values:

$$F(\lambda) = 0.8 F_{ref}(\lambda) \qquad (1b)$$

As stated before, it is widely known that the soft x-rays deposit their energies in the E region at the altitude ranges 100-150 km of the atmosphere. This fact can be verified by using the modified EUVAC spectrum to quantitatively show the distribution of the ionization energies at different altitude ranges. The percentage of ionization energy in the different EUVAC spectral bins at two different altitude ranges are shown as shown in Figure 1a and 1b. In the figures the EUVAC bins below 91.3 nm, which are mainly responsible for ionization, are divided into three groups, the first group is all wavelength bins below 30.4 nm which represent the soft x-rays spectrum, the second group is the He II 30.4nm emission and the last group contains all the ionizing bins above 30.4 nm. For a particular wavelength group, each boxplot represents the distribution of the percentage ionization energy for a range of solar activities (P= 70 -265). The percentage is calculated with respect to the total ionization energy which is taken as the sum of the ionization energy in the solar spectral bins up to 91.3 nm.

Figure 1a shows the ionization energies at an altitude of 120 km which is the E region of the atmosphere. We know that most of the soft x-rays deposit their energies at around this altitude. As seen from the figure, on an average 68% of the input solar energy is in the wavelength bins below 30 nm, i.e., in the soft x-rays, 21% percent is due to 30.4nm emission and the rest 10% is due to the wavelengths above 30.4nm. As we go higher in altitudes, the contribution to the ionization from the soft x-rays begins to diminish as indicated by Figure 1b, which shows the ionization energies in the F region at 300 km. Here most of the ionization, around 50%, occurs due to wavelengths beyond 30.4 nm and the contribution from soft x-rays is about 28% and that from He II is 23% .

Secondly, for any boxplot, the ionization energies for different P values are within 3% of each other. Since the energy in each boxplot does not vary significantly with the F10.7 index, we have considered an F10.7 value of 100 for both daily and 81-day average, for this study.

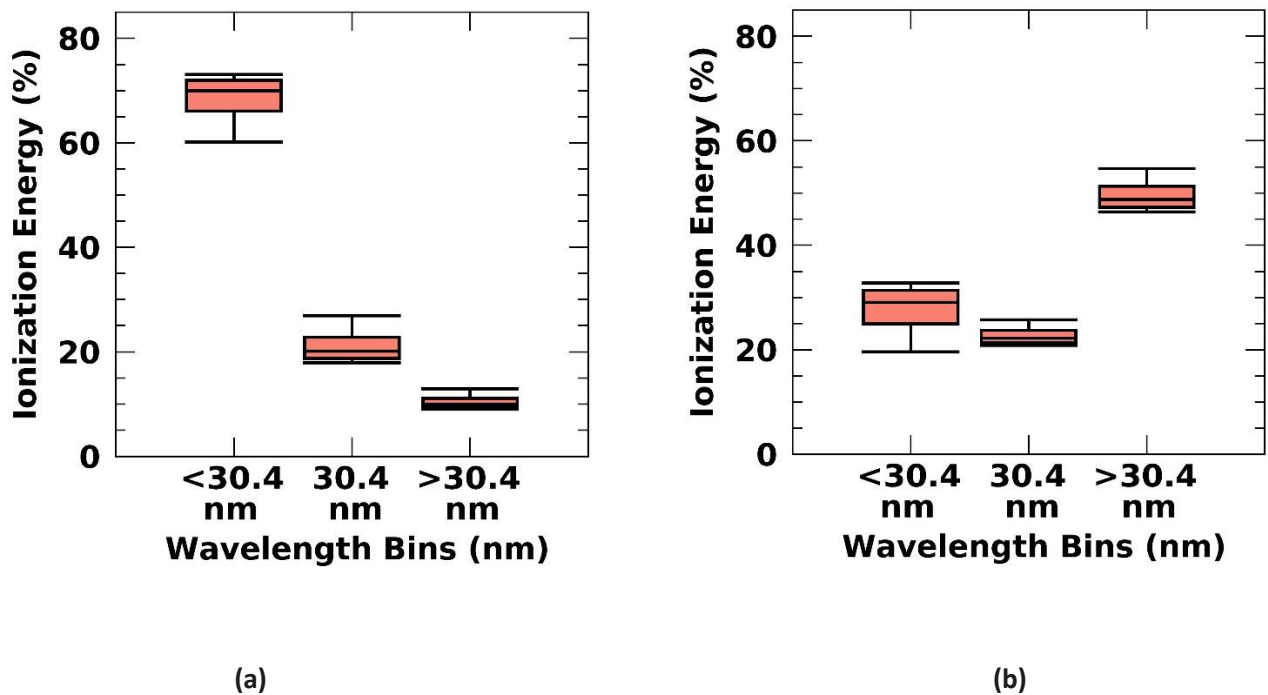

(a)                                                                                          (b)

Figure 1a. Percentage of Ionization Energy in different wavelength ranges at (a) 120 km (E region) (b) 300 km (F region) due to different solar activities (P varied from 70 to 265) . At E region, most of the ionization occurs due to soft x-rays while in the F region, most of the ionization is due to EUV.

## 1.2 Neutral Temperatures

The model uses the one-dimensional time-dependent heat equation given in Equation (2a) (Roble et al. [1987]) to obtain the global mean neutral temperatures. The right-hand side of the heat equation consists of four terms that account for the change in neutral temperature: molecular diffusion, eddy diffusion and neutral heating and cooling rates. Out of these four processes, the eddy diffusion plays an insignificant role in the increase in neutral temperature in the thermosphere above 90 km.

$$\frac{\partial T}{\partial t} = \frac{g}{P_0} \frac{e^Z}{C_p} \frac{\partial}{\partial Z} \left( \frac{K_T}{H} \frac{\partial T}{\partial Z} + K_E H^2 C_p \rho \left( \frac{g}{C_p} + \frac{1}{H} \frac{\partial T}{\partial Z} \right) \right) + \frac{(Q-L)}{C_p} \tag{2a}$$

Here T is the neutral gas temperature, t is the time, $C_p$ is the specific heat per unit mass at constant pressure, H is the scale height, $K_T$ and $K_E$ are the thermal and eddy thermal conductivity coefficients respectively and $\rho$ is the average mass density, Q and L are the globally averaged heating and cooling rates, $P_0$ and g are the model reference pressure (50 μu Pa) and the acceleration due to gravity respectively.

The specific heat capacity and thermal conductivity are defined in Banks and Kockarts [1973] as:

$$C_p = k \left( \frac{7}{2} \left( \frac{v_{N_2}}{m_{N_2}} + \frac{v_{O_2}}{m_{O_2}} \right) \right) + \frac{5}{2} \frac{v_O}{m_O} \tag{2b}$$

$$K_T = \left( 56(v_{O_2} + v_{N_2}) + 75.9 v_O \right) T^{0.69} \tag{2c}$$

where $v_i$ and $m_i$ refer to the volume mixing ratio and molecular or atomic masses respectively of the species i and K is the Boltzmann's constant.

The eddy diffusion term (Dickinson et al [1984]) assumed an a globally averaged eddy diffusion coefficient given by:

$$K_E(Z) = 5 \times 10^{-6} exp(-7 - Z) s^{-1} \tag{2d}$$

Where Z is in log pressure units defined in Equation 2e

$$Z = -ln \left( \frac{P}{P_0} \right) \tag{2e}$$

where p is the pressure at the grid point Z and $P_0$ is the reference pressure of 50uPa. A one unit change in Z corresponds to a change in altitude by one pressure scale height or e change in pressure. The upper and lower vertical boundaries of the model is given by Z= 7 and Z=-7 respectively, with a resolution of $\Delta$ Z=0.25. The lower boundary corresponds to an altitude of 97 km as calculated by Roble et al. [1987] and corresponds to mesopause boundary averaged over one solar cycle. The upper boundary depends on the solar activity and exospheric temperature and typically varies between 450 km to 750 km.

The expression in Equation 2d can be converted into dimensional units by multiplying it with the square of the scale height and has a value of 150 m² s⁻¹ at Z=-7. Equation 2d is exponentially decreasing with Z so as to confine the effects of eddy diffusion to the lowest altitudes of about 97 km (Dickinson et al. [1984]).

The temperature at the lower boundary of the thermosphere in the model is obtained from NRLMSIS (Picone et al. [2002]) and is dependent on solar activity, while at the upper boundary thermal conduction is the main heating process and therefore, we assume $\frac{\partial T}{\partial Z} = 0$.

## 1.3 Thermospheric Heating Processes

The solar XUV radiation ionizes the neutrals and produces ions and free electrons. Further collisions of neutrals with ions and electrons, which increases their kinetic energy, leads to an increase in the neutral temperature. Different processes contribute to the local heating of the neutral thermosphere at different altitudes. These processes are shown in Figure 2.

The first two processes are the heating due to the absorption of UV radiation in the Shumann Runge (SR) band (175-205 nm) and SR continuum (125-175 nm) by $O_2$. The absorption of solar radiation in the SR bands and continuum by $O_2$ leads to the dissociation of the molecule and thus transfer of the heat energy (photon energy minus the dissociation energy) to the O atoms. These are important heating processes in the lower altitude ranges of 60 to 120 km approximately. The ACE1D model uses a parameterized expression for heating due to absorption of $O_2$ in the SR bands obtained from Strobel et al. [1978]. The heating rate due to the SR continuum is obtained by a more detailed analytic expression from DeMajistre et al. [2001]

The complete list of exothermic neutral-neutral reactions, ion-neutral and ion-recombination reactions have been included in the model. These reactions absorb the EUV radiation and heat the neutral gas by the excess energy released during the reaction. A significant source of neutral heating in the lower thermosphere at about 97km is the recombination of atomic oxygen. Quenching of excited atomic oxygen ($O(^1D)$) by neutral species ($N_2$, $O_2$ and O) is important above 110 km. Finally, the exothermic reactions of $N(^2D)$ and $N(^4S)$ with $O_2$ to form NO at higher vibrationally excited state ($v \geq 1$) also lead to heating of the neutral thermosphere at 150 km, however the effectiveness of the process is reduced by the chemiluminescence of (NO)in which the molecule is returned to its lower state of vibrational excitation (v' ≤ v)) with the loss of energy as radiation (Venkataramani [2018]).

In order to make the global average temperature agree with the results from the NRLMSIS model (Picone et al [2002]), a 70 GW joule heating as suggested by Foster et al. [1983] for geomagnetically quiet conditions has been used in the ACE1D model. Joule heating can be expressed by Equation 3 as:

$$Q_{Joule} = \sigma_p E^2 \qquad (3)$$

where $\sigma_p$ is the Pederson Conductivity, E is the superimposed electric field, varying between 6.47-4.07 mV m$^{-1}$. Joule heating is partitioned between neutrals and ions. However, the ions rapidly transfer their energy to the neutrals through collisions, therefore, the entire Joule heating given by Equation 3 is used to heat the neutral thermosphere.

Most of the energy of the photoelectrons are either used to further ionize or excite neutrals or heat ambient electrons. The heating rate of neutrals due to direct collisions with photoelectrons is therefore small and is assumed to be 5\% of the total EUV energy absorbed (Stolarski [1976]; Roble et al [1987]).

The mid and upper thermosphere, there are several ion species. Therefore, an important neutral heating at the upper thermosphere is due to the collisions of ambient or thermal electrons, ions and neutrals (Schunk [1988]). At higher altitude, an important mechanism of plasma transport is due to the ambipolar diffusion due to temperature variation. The temperature gradient in the thermosphere causes the ions and electrons to move from the lower to the higher altitudes. However due to the smaller mass of electrons, they can move faster than the ions, causing a charge separation. This sets up a polarizing electric field which prevents further charge separation. Once this field is set up, the ions and electrons move together as a single gas under the influence of the aforementioned temperature gradient and upon collision with the neutral gas particles results in heating of the neutral thermosphere.

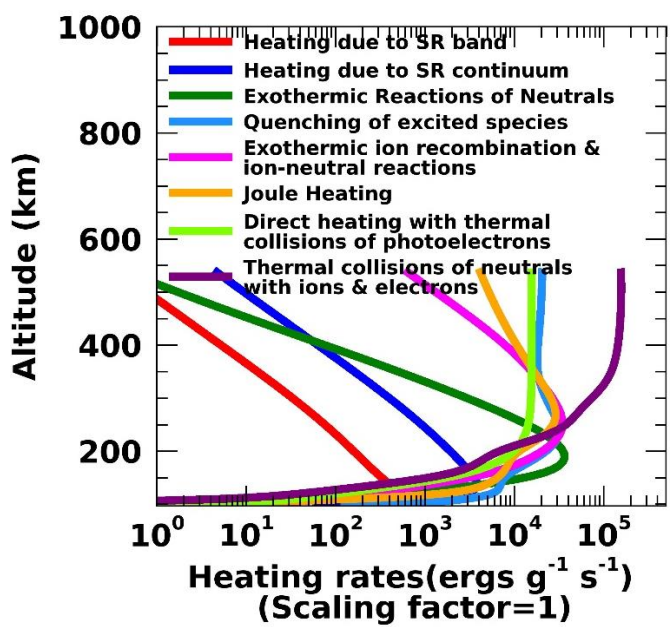

**Fig 2: Heating (in logarithmic scale ) due to absorption in the Schumann Runge bands and Schumann Runge continuum; exothermic reactions of neutral species; quenching of excited species; direct heating due to thermal collisions with photoelectrons; exothermic ion recombination and ion-neutral reactions; joule heating and thermal collisions of neutrals with ions and electrons**

In the next few sections, we conduct a numerical experiment on the model to show that the exospheric temperature is most sensitive to changes in the irradiance in the solar soft x-ray wavelengths. Next, we multiply the input solar soft x-rays flux by different scaling factors to emulate the inherent uncertainty in the data and analyze the changes in the neutral temperatures from approximately 100 km to 600 km. Then we analyze each of the terms of the thermodynamic heat equation used in the ACE1D model to explain the process by which these soft x-rays heat the entire thermosphere. Lastly, we show the relation between the uncertainty in our modeled exospheric temperature and solar activity as indicated by the F10.7 index.

 2. Numerical Experiment

The study of the effect of each solar spectral bin on the modeled neutral temperature is inspired by conducting a numerical experiment on the 37 bins of the EUVAC input spectrum.  The change in exospheric temperature per unit wavelength for each wavelength bin is generated by running the ACE1D model iteratively, with the solar flux in one wavelength bin set to zero in each case. The changes in exospheric temperatures are obtained by calculating the difference of these temperatures from a reference temperature. The reference temperature is calculated without any modifications to the solar spectrum. To account for the variable wavelength bin sizes in the EUVAC spectrum, we divide the change in the exospheric temperature in each bin by the width of the corresponding bin. The resulting percentage change in the exospheric temperature per nm is shown in Figure 3. From the figure, it can be seen that the maximum change in exospheric temperature in the soft x-ray region is between 10 nm and 30.4 nm and is about 0.94%.  The change in exospheric temperature for the strong He II emission at 30.4 nm is 2.8%. Above this wavelength, the significant changes in the exospheric temperatures are due to the He I 58.4 nm (0.32 %) and H Lyman- beta 102.6 nm (0.04 %).

This experiment proves that the exospheric temperatures are sensitive to changes in irradiance in the soft x-ray spectrum. It is also sensitive to the strong He II 30.4 nm emission as seen from the figure. In fact, it is shown in the following sections that the contribution of the solar soft x-rays irradiance to the change in the neutral temperature is comparable to that of the He II 30.4 nm emission.

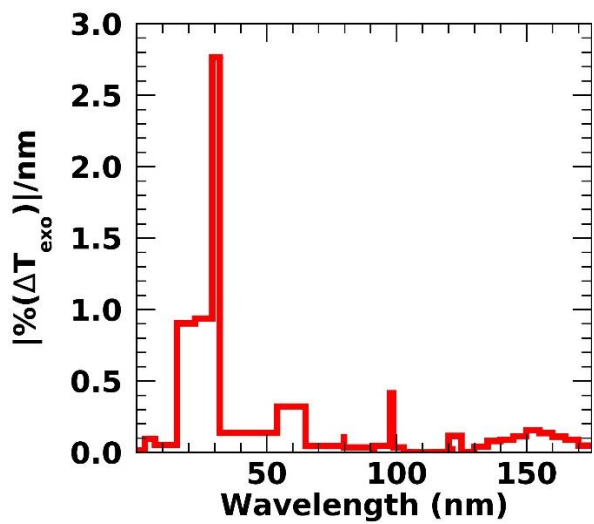

Figure 3: Change in Exospheric temperature of the neutral atmosphere per wavelength bin as a function of wavelength. Maximum change in temperature occurs in the soft x-ray range

3. Results

Uncertainty in the solar soft x-rays irradiance is simulated by scaling only the solar flux in the soft x-rays wavelength bins in the model by different scaling factors and then studying the changes in neutral temperatures. There are eight spectral bins in the EUVAC model which represent the soft x-rays: 0.05 nm-0.4 nm, 0.4 nm-0.8 nm, 0.8 nm-1.8 nm, 1.8 nm-3.2 nm, 3.2 nm-7 nm, 7 nm-15.5 nm, 15.5 nm-22.4 nm and 22.4 nm-29 nm). We have used a total of five scaling or multiplicative factors (zero, one half, one, two and four) to scale these eight spectral bins, while keeping the rest of the bins unchanged. A scaling factor of zero implies that the soft x-rays are absent, a scaling factor of one implies that the input solar spectrum is unchanged and scaling factors of two and four implies the soft x-rays irradiance is increased twice and four times, respectively. The ACE1D model is used with these five solar spectra as inputs and an F10.7 of 100 to generate the neutral temperatures as a function of altitude. Other parameters in the neutral temperature analyzed in the following sections are consequently scaled by these five scaling factors.

The same procedure, as described above, is followed for the analysis of heating of the upper thermosphere due to the He II emission.

### 3.1 Results and Inference of scaling of the soft x-rays below 30.4 nm

Figures 4a and 4b show the plots of the neutral temperature and ionization energy respectively as functions of the altitude for the different scaling factors. The altitude of peak deposition of ionization energy from the solar radiation is the lower thermosphere at about 120 km as seen in figure 4b. With the increase in soft x-ray flux, the ionization also increases. However, as seen in figure 4a, the increase in neutral temperature with the solar soft x-rays is more pronounced in the upper thermosphere, above the altitude of the maximum energy deposition.

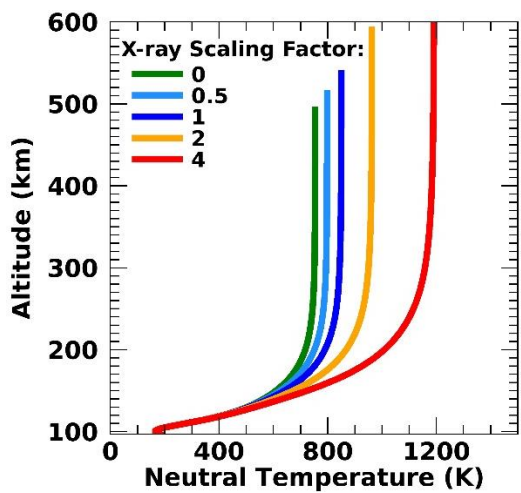 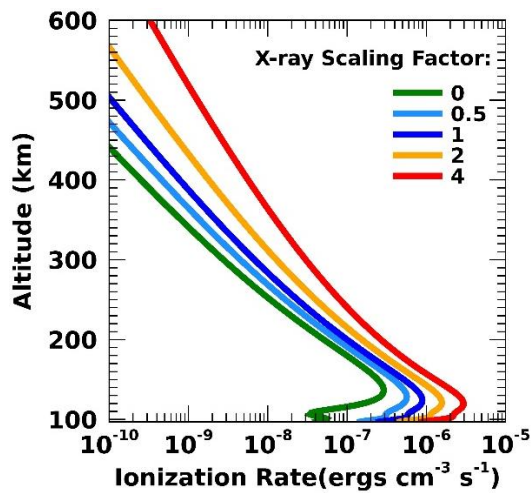

(a) (b)

**Figure 4 (a) Neutral Temperature and (b) Ionization Energy as functions of Altitude. The different scaling factors scale the Soft x-rays below 30.4 nm. With increase in soft x-ray scaling, the neutral temperature and ionization energy also increase**

We use the thermodynamic heat equation as described in section 2 to attribute this increase in temperature to the thermal conduction of the energy of the soft x-rays from the lower to upper thermosphere. To prove this we show the different terms of the Equation 3a in Figure 5.

Figure 5a shows the thermal conduction of heat as a function of different altitudes for the different soft x-rays scaling factors. With the increase in the soft x-rays irradiance the thermal conduction is seen to increase, and this effect is pronounced at higher altitudes starting at about 200 km. In fact, compared to the heating and cooling rates shown in Figures 5b and c respectively, the thermal conductivity is almost five times more than the heating rate and almost ten times more than the cooling rate at an altitude of 500 km. Therefore, thermal heat conduction is the dominant process of heat transfer at the upper thermosphere.

The Eddy Diffusion, shown in Figure 5c, is a turbulent mixing process and it is significant only at altitudes below 90 km where the atmosphere is well mixed and has an exceedingly small contribution at higher altitudes. This is shown in the figure where its contribution to the heat equation at lower altitudes is very small and becomes negligible at higher altitudes.

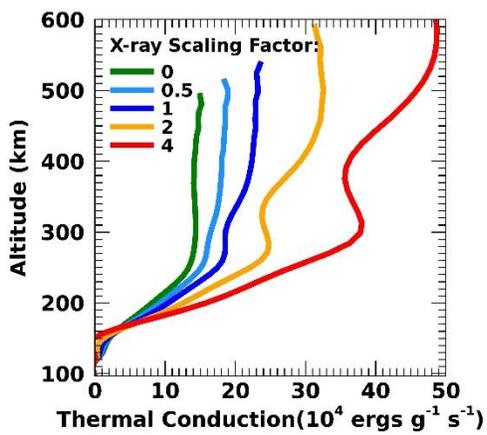
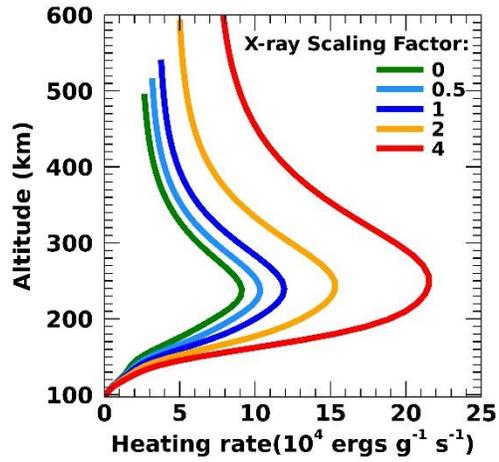

(a)            (b)

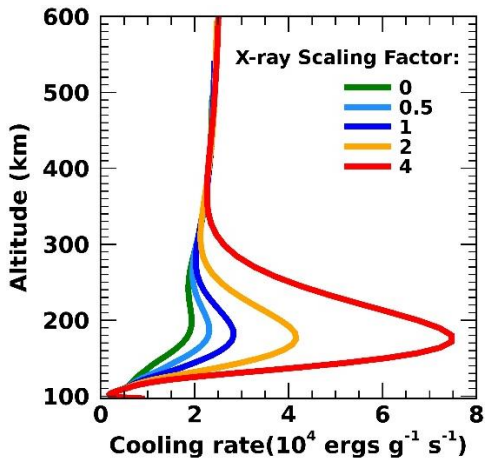
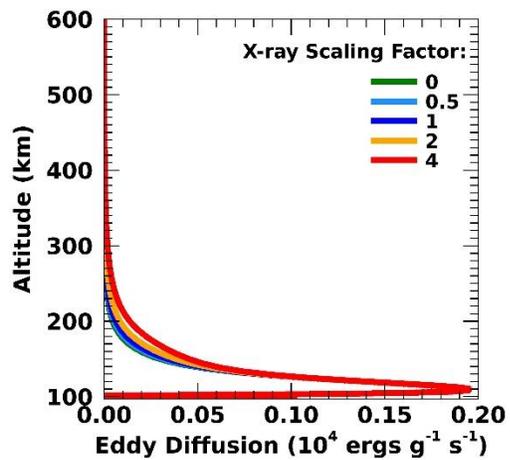

(c)            (d)

**Figure 5(a) Thermal Conduction (b) Heating and (c) Cooling rates (c) Eddy diffusion as functions of altitude. The different scaling factors scale the Soft x-rays below 30.4 nm. With increase in soft x-ray scaling, the thermal**

**conduction and heating rate increase at high altitude and the cooling rate increase at low altitude. The effect of eddy diffusion is negligible at high altitude.**

Figure 6 shows the different heating processes in the thermospheric, in linear scales that are included in the ACE1D model. There are eight main heating processes, namely heating due to the absorption of the solar radiation by $O_2$ in the Schumann Runge bands and continuum; exothermic reactions of neutral species; quenching of excited species; direct heating due to thermal collisions with photoelectrons; exothermic ion recombination and ion-neutral reactions; Joule heating; and thermal collisions of neutrals with ions and electrons. From the linear plot, it is apparent that at high altitudes near 500 km, heating due to the thermal collisions of neutrals with the ions and electrons is greater than all the other heating mechanisms by at approximately one order of magnitude, therefore making it the main contributing factor to the heating of the upper thermosphere.

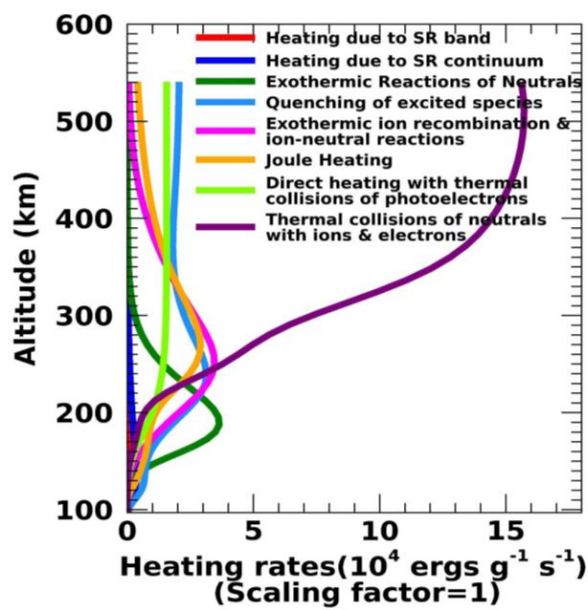

**Fig 6: Heating (in linear scale) due to absorption in the Schumann Runge bands and Schumann Runge continuum; exothermic reactions of neutral species; quenching of excited species; direct heating due to thermal collisions with**

**photoelectrons; exothermic ion recombination and ion-neutral reactions; joule heating and thermal collisions of neutrals with ions and electrons. The soft x-rays are not scaled (Scaling factor =1) in obtaining these rates.**

Figure 7a shows the electron density profile as a function of altitude. With the increase in soft x-rays irradiance, the electron densities also increase at all altitudes due to the increase in ionization. The total heating by thermal collisions of neutrals with ions and electrons are separated and shown individually in Figure 7b. Both the heating by ions and electrons increases with increase in soft x-rays irradiance. Furthermore, the heating by ions increases with increase in altitudes while the maximum electron heating is achieved at a lower height of about 200 km. At higher altitudes, the ionic heating is two orders of magnitude greater than the electronic heating.

Therefore, the ionization energy that is deposited by the soft x-rays at the lower altitude is carried by the collisions of neutrals with ions and electrons to the higher altitudes.

This is further proven by Figure 7c, which shows the ion and electron temperatures. At higher altitudes, we see that the electron temperatures decrease with the increase in the soft x-rays flux. The collisions of electrons with the neutrals cause an increase in the neutral temperature and a subsequent decrease in the electron temperatures at the exosphere. The change in the ion temperatures is not that appreciable at higher altitudes.

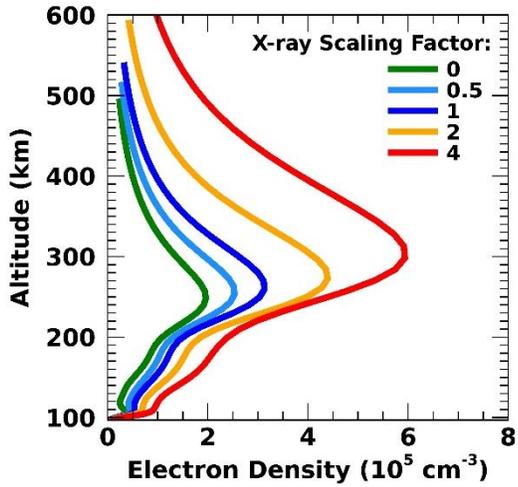

(a)

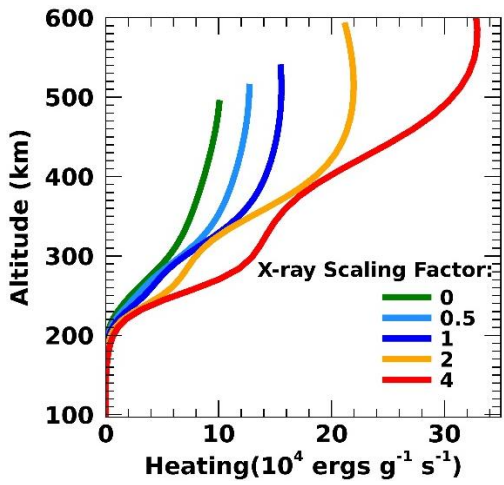

(b)

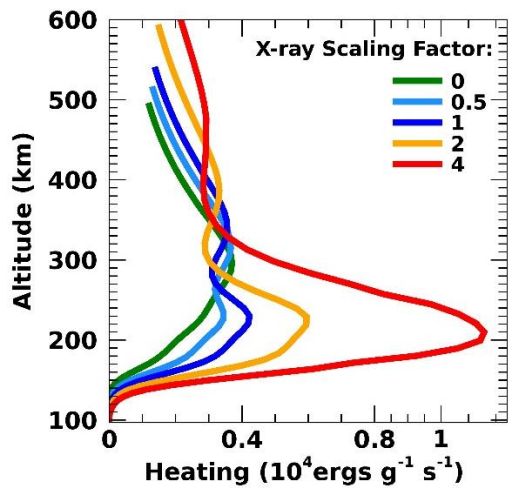

(c)

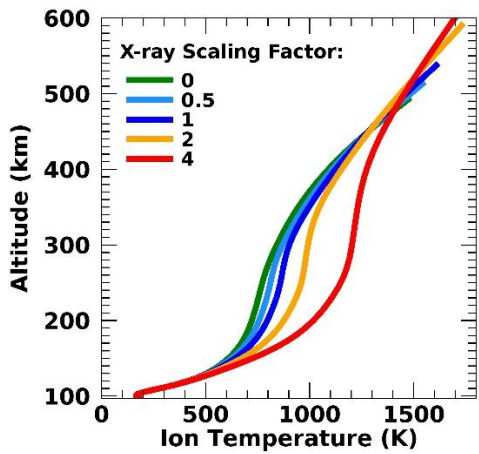

(d)

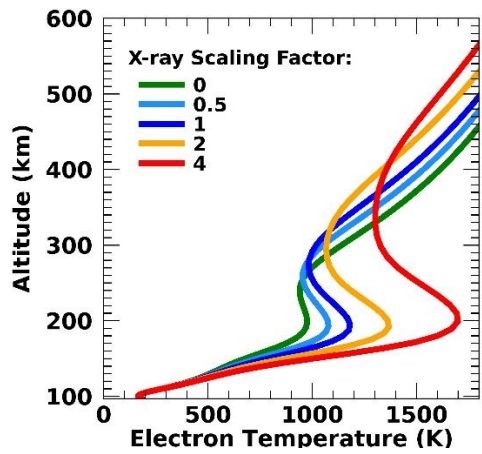

(e)

**Figure 7(a) Electron density (b) Heating of neutrals by ions and (c) electrons.(d) Ion and (e) Electron Temperatures as functions of altitude. The different scaling factors scale the Soft x-rays below 30.4 nm. With increase in soft x-ray scaling, the electron density, heating of neutrals by ions and electrons and ion temperature increase. The electron temperature decrease with increase in soft x-rays irradiance as their energy is transferred to the neutral atmosphere.**

### 3.1.1. Uncertainty in Exospheric Temperature for different F10.7 for scaling of x-rays below 30.4nm

Figure 8 shows the uncertainty in the exospheric temperature with the F10.7 index. Since the F10.7 index is a measure of the solar activity, Figure 8 is the estimation of the uncertainty in the exospheric temperature with the change in solar activity. The uncertainty is calculated in the following way: For a particular F10.7, we calculate three exospheric temperatures corresponding to three scaling factors of 0.5, 1 and 2 of the input solar irradiance bins. The three scaling factors represent the uncertainty in the measurement of the solar flux in the soft x-rays region of the spectrum. The exospheric temperature, with a scaling factor of 1, is taken as the reference temperature. We then calculate the uncertainty of the two temperatures from the reference exospheric temperature. We repeat this process for different F10.7 starting from 50 up to F10.7 of 250. From the figure, we see that the uncertainty increases almost linearly with the increase in the F10.7. The maximum uncertainty is about 260 K at a F10.7 of 250.

Since the change in exospheric temperature is linear with the change in F10.7 and we have calculated the uncertainty by changing the soft x-rays by a scaling factor of 2, therefore, based on the above results, we estimate the impact of a factor-of-two uncertainty in the solar soft X-ray irradiance on thermospheric temperatures. It is well known that the thermospheric temperature strongly depends on heating due to the solar EUV radiation, and since the uncertainty in this study is calculated on the exospheric temperature, therefore, this result shows that the solar soft x-rays have a significant effect on the global thermospheric temperature as well.

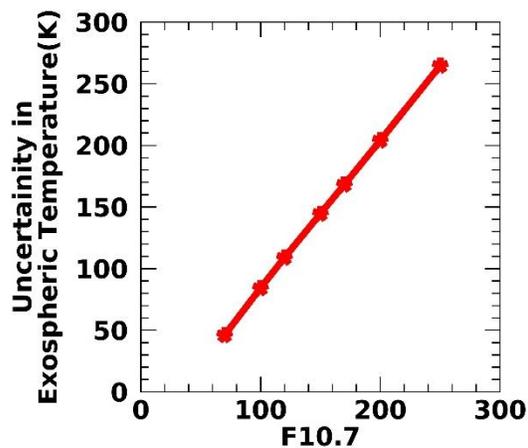

**Fig 8: Uncertainty in Exospheric Temperature for different F10.7 for scaling of soft-rays below 30.4nm. Uncertainty increases with the increase in solar activity**

**3.2 Scaling of the HeII 30.4 nm solar emission**

In this section, we compare the effects of change in the thermospheric temperature (with change in the solar irradiance) due to He II emission with that of the soft x-rays analysis of Section 3.1. The analysis of the heating of the thermosphere, carried out in section 3.1, is repeated in this section in the subsequent figures but with only the solar flux of the 30.4 nm wavelength, i.e. He II emission, scaled by the five scaling factors, while keeping the rest of the bins unchanged. There is only one wavelength bin, from 29 nm to 32 nm, in the input solar spectrum of the model, which represents this emission line.

The He II line is a strong solar emission, so it is important to analyze its role in the thermospheric heating. Figure 9a shows the neutral temperature as a function of altitude for the five different scaling factors, following the method applied in the previous section. The neutral temperatures calculated in the previous section, by changing the soft x-rays intensities, are approximately 1.2 times larger than that obtained here (from scaling of the He II emission). Therefore, we see that the effects of the solar soft x-rays on the neutral exospheric temperature is comparable to that of this strong 30.4 nm solar emission.

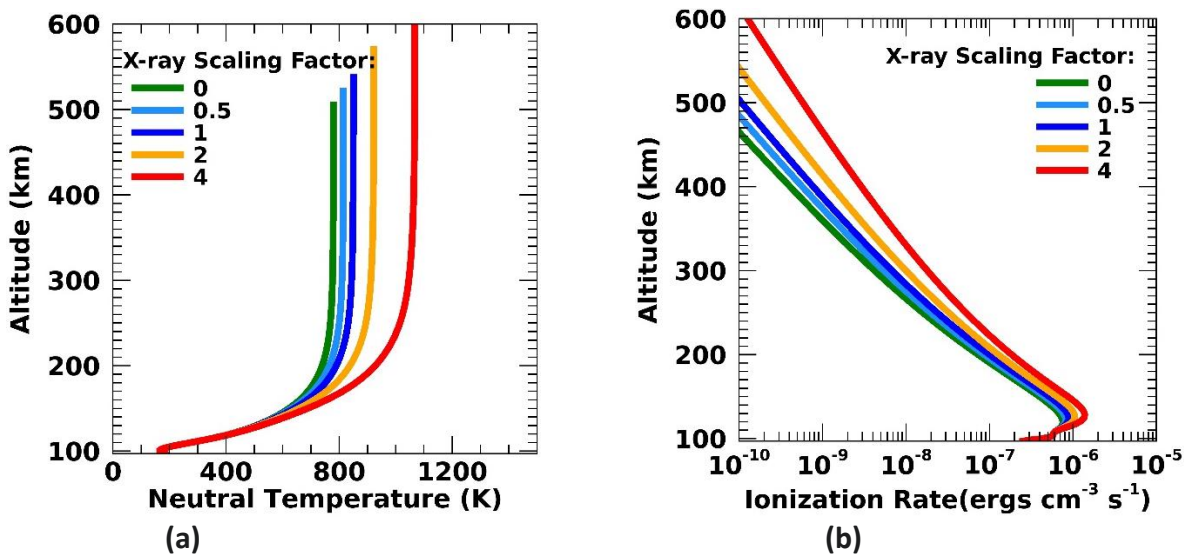

**Figure 9 (a) Neutral Temperature and (b) Ionization Energy as functions of Altitude. The different scaling factors scale the He II 30.4 nm emission. With increase in He II scaling, the neutral temperature and ionization energy also increase**

The mechanism of increase in the thermospheric temperature can be analyzed as was done in Section 3.1. The different processes affecting the temperature change, i.e., the thermal conduction, heating and cooling rates and eddy diffusion, calculated from the thermodynamic heat equation is shown in figure 10. As in the previous case, the eddy diffusion is negligible compared to the other processes in the altitude range of interest and thus can be ignored. By following the previous analysis, we can see that in the case of He II emission, the main contribution to the change in thermospheric temperature at higher altitudes is due to the heat conduction by non-thermal electrons.

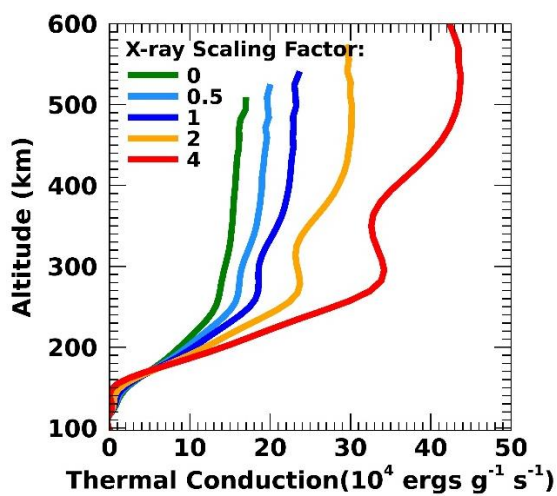

(a)

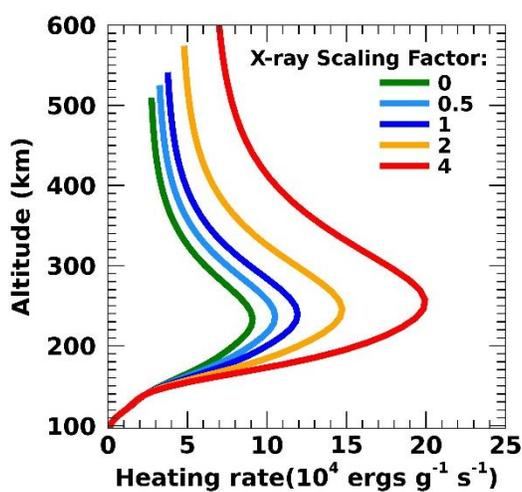

(b)

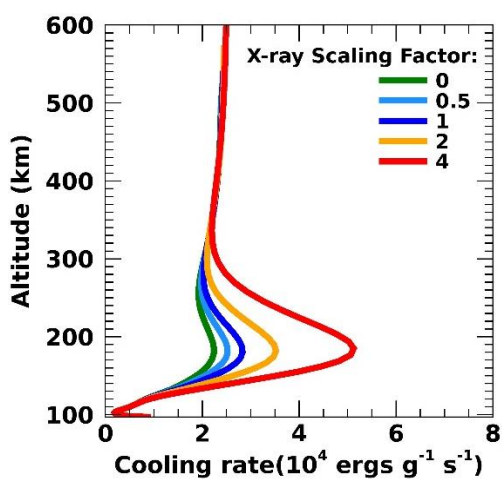

(c)

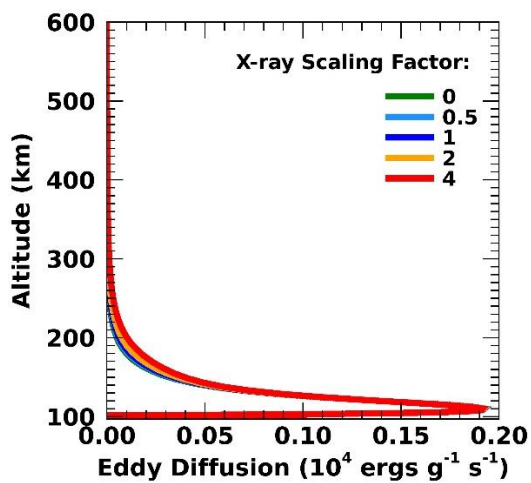

(d)

**Figure 10(a) Thermal Conduction (b) Heating and (c) Cooling rates (c) Eddy diffusion as functions of altitude. The different scaling factors are applied to the He II 30.4 nm emission. With increase in He II irradiance, the thermal conduction and heating rate increase at high altitude and the cooling rate increase at low altitude. The effect of eddy diffusion is negligible at high altitude.**

Figure 11a shows the electron density profile scaled in response to the scaling of the soft x-rays irradiance. Figures 11 b and c show the heating of the neutrals by electrons and ions, and the ion and electron temperatures respectively.

Due to the thermal gradient set up by the deposition of ionization energy at the lower altitudes, the electrons with their higher thermal velocities and lower mass move up in the thermosphere quicker than the ions, which set up the polarization electric field. Once equilibrium is reached the plasma moving through the neutral atmosphere via ambipolar diffusion will heat it, causing fall of electron temperatures.

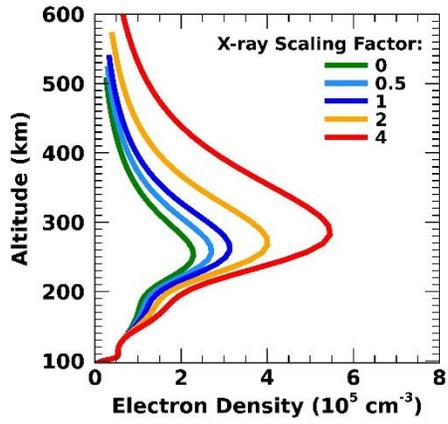

(a)

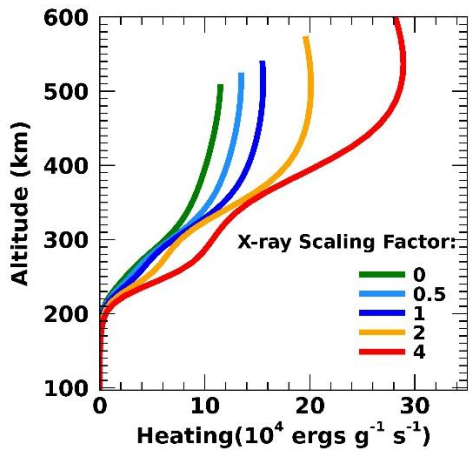

(b)

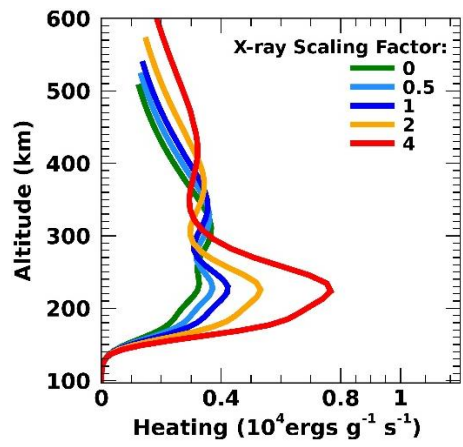

(c)

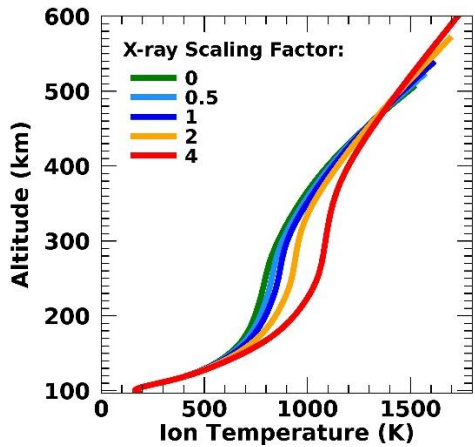

(d)

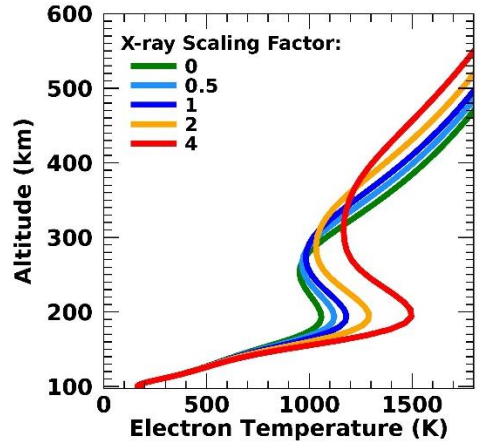

(e)

Figure 11(a) Electron density (b) Heating of neutrals by ions and (c) electrons.(d) Ion and (e) Electron Temperatures as functions of altitude. The different scaling factors scale the He II 30.4 nm emission. With increase in He II scaling, the electron density, heating of neutrals by ions and electrons and ion temperature increase. The electron temperature decrease with increase in soft x-rays irradiance as their energy is transferred to the neutral atmosphere.

### 3.2.1. Uncertainty in Exospheric Temperature for different F10.7 for scaling of x-rays below 30.4nm

Figure 12 shows the uncertainty in the exospheric temperature for different F10.7. The method of calculating the uncertainty has already been discussed in section 3.1. For scaling of only the He II emission by factors of 0.5,1 and 2, the uncertainty increases almost linearly with the increase in F10.7 and has a maximum uncertainty of approximately 105 K at a F10.7 of 250, as compared to the scaling of the soft x-rays which has a maximum uncertainty of 250K at a F10.7 of 250 . This is reasonable because the soft x-rays cover a larger energy range and consist of very high energy solar flux with varying intensities than the very narrow He II emission, which is better represented by the input EUVAC solar spectrum.

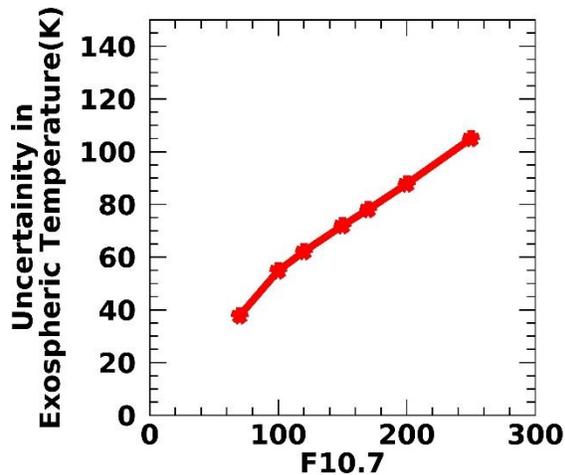

**Fig 12: Uncertainty in Exospheric Temperature for different F10.7 for scaling of He II 30.4nm emission. Uncertainty increases with the increase in solar activity**

4. **Comparisons With GUVI Temperature Data**

   The dayside exospheric neutral temperature obtained from the Global Ultraviolet Imager (GUVI) onboard the Thermosphere-Ionosphere-Mesosphere Energetics and Dynamics (TIMED) satellite (Meier et al [2015]) has been compared to the modelled NRLMSIS dayside and global mean temperature and also the global average exospheric temperature obtained from the ACE1D model. The four temperatures are shown in Figure 13 as a function of F10.7. GUVI daily temperatures were obtained from a period of 2002 till 2007 and were averaged over the entire dayside global area to obtain the average dayside temperature. The ACE1D global mean temperature as a function of F10.7 is on an average 0.95 times that of the NRLMSIS global mean. The daytime average neutral temperature obtained from the NRLMSIS model agrees very well with the GUVI daytime temperature data. Only at F10.7 less than 90, i.e., for low solar activity the data and model temperatures converge for daytime and global average values. Overall, the daytime average exospheric is approximately 1.3 times the global (day and night) average temperature.

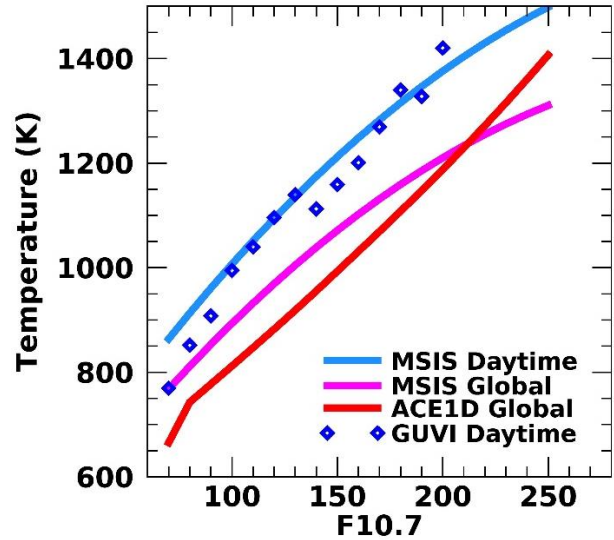

**Figure 13: Comparisons of GUVI, ACE1D and MSIS average Exospheric Temperatures. The GUVI Daytime average temperature agrees with the MSIS daytime temperature. The ACE1D Global average temperature are closer to the the MSIS Global average temperature.**

## 5. Conclusion

The dependence of global thermospheric temperature on the solar soft x-rays flux below the He II 30.4 nm emission, is studied here using a newly developed one-dimensional ACE1D model. We show that the solar soft x-rays below 30.4 nm are important drivers of the temperature of the entire thermosphere and ionosphere, although most of their ionization energy is deposited in the lower thermosphere. The solar soft x-rays flux, which is the most variable part in the solar spectrum, is represented by eight bins varying from 0.05 nm till just below 30.4 nm in a modified input spectrum (EUVAC) consisting of 37 bins from 0.05 nm – 175 nm and F10.7 value.

In order to study the effects of the uncertainty in the solar soft x-rays irradiance, we use five different scaling factors to scale the soft x-rays flux and observe their effects on the thermospheric temperature. We then utilize the thermodynamic heat equation to conclude that the main source of this temperature increase in the upper thermosphere is due to the conduction of heat from the lower to the upper thermosphere via collisions of neutrals with ions and electrons.

We further scaled the HeII 30.4 nm solar flux in the input spectrum and compared its effects with that of the soft x-rays scaling. We found that the effect of the varying the soft x-rays irradiance on the neutral temperatures is about 1.2 times more than that of the He II emission, which implies that the soft x-rays are significant drivers of the thermospheric temperature and are as important as the strong HeII emission. Moreover, the uncertainty in the calculation of the neutral temperatures due to the variation of the solar flux in the soft x-rays wavelengths leads to errors in the model-predicted thermospheric temperature and this uncertainty is almost linear with the increase in F10.7. Hence, there is a need for better models that can reflect this variability in the solar soft x-rays flux.